\documentclass{aa}
\usepackage{graphicx}
\usepackage{rotating}
\usepackage{txfonts}
\usepackage{comment}

\usepackage{xcolor}
\usepackage{ulem}

\newcommand{\hi}{H\,\textsc{i}}
\newcommand{\hii}{H{\sc i}\,21cm}
\newcommand{\apx}{$\sim$ }
\newcommand{\pc}{$\%$ }

\newcommand{\lt}{$<$ }
\newcommand{\gt}{$>$ }

\newcommand{\target}{\object{3C\,433}}
\newcommand{\eg}[1]{\citep[e.g.][]{#1}}
\newcommand{\kmps}{km~s$^{-1}$}
\newcommand{\p}[1]{$^{-#1}$}
\newcommand{\pp}[1]{$^{#1}$}

\newcommand{\plm}{$\pm$}
\newcommand{\tim}{$\times$}
\newcommand{\halpha}{H$\alpha$}

\newcommand{\beq}{\begin{equation}}
\newcommand{\eeq}{\end{equation}}

\newcommand{\tspin}{{T$_{\rm spin}$}}

\newcommand{\Mstar}{$\textrm{M}_*$}
\newcommand{\Msun}{M$_\odot$}
\newcommand{\MHI}{M$_\textrm{HI}$}

\begin{document}

   \title{Disc galaxy resolved in \hi\ absorption against \\the radio lobe of \target: Case study for future surveys}

   \author{Suma Murthy\inst{1,2},
          Raffaella Morganti\inst{2,1},
          Bjorn Emonts\inst{3},
         Montserrat Villar-Mart\'in\inst{4},
         Tom Oosterloo\inst{2,1},
         Reynier Peletier\inst{1}
         }

   \institute{Kapteyn Astronomical Institute, University of Groningen, Postbus 800, 9700 AV Groningen, The Netherlands
             \email{murthy@astro.rug.nl}
         \and
             ASTRON, the Netherlands Institute for Radio Astronomy, Postbus 2, 7990 AA, Dwingeloo, The Netherlands
         \and         
            National Radio Astronomy Observatory, 520 Edgemont Road, Charlottesville, VA 22903, USA
         \and
             Centro de Astrobiolog\'ia, (CAB, CSIC-INTA), Departamento de Astrof\'isica, Cra. de Ajalvir Km. 4, 28850$-$ Torrej\'on de Ardoz, Madrid, Spain}

   \date{Received August 5, 2020, accepted September 17, 2020}

 
   \abstract{The neutral atomic gas content of galaxies is usually studied in the \hi\ 21cm emission line of hydrogen. However, as we go to higher redshifts, owing to the weak strength of the transition, we need very deep integrations to detect \hi\ emission. The \hi\ absorption does not suffer from this dependence on distance as long as there is a sufficiently bright radio source to provide the background continuum. However, resolved \hi\ absorption studies of galaxies are rare. We report one such rare study of resolved \hi\ absorption against the radio galaxy \target\ at $z = 0.101$, detected with the Very Large Array (VLA). The absorption was known from single-dish observations, but owing to the higher spatial resolution of our data, we find that the absorber is located against the southern lobe of the radio galaxy. The resolved kinematics shows that the absorber has regular kinematics with an \hi\ mass $\lesssim$ 3.4 $\times$ 10\pp{8} \Msun\ for \tspin $=$ 100K. We also present deep optical continuum observations and \halpha\ observations from the Gran Telescopio CANARIAS (GTC), which reveal that the absorber is likely to be a faint disc galaxy in the same environment as \target, with a stellar mass of \apx 10\pp{10} \Msun\ and a star-formation rate of 0.15~\Msun~yr\p{1} or less. Considering its \hi\ mass, \hi\ column density, stellar mass, and star-formation rate, this galaxy lies  well  below  the  main  sequence  of  star  forming galaxies. Its \hi\ mass is lower than the galaxies studied in \hi\ emission at $z \sim 0.1$. Our GTC imaging has revealed, furthermore, interesting alignments between H$\alpha$ and radio synchrotron emission in the \hi\ companion and in the host galaxy of the active galactic nucleus as well as in the circumgalactic medium in between. This suggests that the shock ionization of gas by the propagating radio source may happen across a scale spanning many tens of kpc. Overall, our work supports the potential of studying the \hi\ content in galaxies via absorption in the case of a fortuitous alignment with an extended radio continuum source. This approach may allow us to trace galaxies with low \hi\ masses which would otherwise be missed by deep \hi\ emission surveys. In conjunction with the deep all-sky optical surveys, the current and forthcoming blind \hi\ surveys with the Square Kilometre Array (SKA) pathfinder facilities will be able to detect many such systems, though they may not be able to resolve the \hi\ absorption spatially. Phase 1 of the SKA, with its sub-arcsecond resolution and high sensitivity, will be all the more able to resolve the absorption in such systems.}


\titlerunning{Resolved HI absorption towards 3C\,433}
\authorrunning{Murthy et al.}
\maketitle

\section{Introduction} \label{introduction}

Neutral atomic hydrogen (\hi) plays an important role in the formation and evolution of galaxies. Hence, attaining a deeper understanding of this component of gas in galaxies (i.e. the content, distribution, kinematics, etc.) is of interest in the study of galaxy evolution. The \hi\ content, traced by the 21cm line, in galaxies in the local Universe has been studied extensively in emission \eg{Wright71, Rogstad74, Sancisi76, vanderkruit78, vanderhulst79, vanderhulst87, Oosterloo93, Barnes01, Koribalski04, Giovanelli05a, Giovanelli05b, Catinella10, Catinella12, Spekkens14, Odekon16, Jones18}. These studies have provided insights into the relation between \hi\ content and stellar mass, galaxy structure, star-formation rate, etc. There have also been extensions of such studies beyond the local universe \eg{Catinella08, Catinella15, Verheijen07, Hess19, Bera19, Bluebird20, Gogate20}. However, even with very deep integrations, direct detections of \hi\ in emission have so far been from gas-rich, star-forming galaxies \citep[$\geq$ 2 $\times$ 10\pp{9}\Msun; e.g.][]{Zwaan01, Catinella12, Fernandez16, Hess19} and mostly for redshifts below $z\sim 0.1$.

\hi\ absorption studies do not suffer from this limitation of redshift because the detection of \hi\ in this case is possible within short integration times provided there is a strong radio continuum in the background. Hence, \hi\ absorption studies have been employed extensively to study atomic gas in damped Lyman-$\alpha$ systems and MgII absorbers \eg{Kanekar09, Gupta09, Kanekar14} and the host galaxies of active galactic nuclei \citep[AGN; e.g.][and references therein]{Morganti05b, Morganti13, Allison15, Aditya16, Aditya17, Chowdhury20, Morganti18} as well as in the study of the distribution of \hi\ in the halo of galaxies \eg{Gupta10, Borthakur16, Dutta17c}. However, by design, these studies have mostly focused on compact radio sources.

In cases where the absorption is detected against an extended radio continuum, it is possible to study the absorber in greater detail. The properties of \hi\ at parsec scales have been probed against the extended radio continuum using very long baseline interferometry techniques \citep[e.g. ][]{Borthakur10, Srianand13, Gupta18, Schulz18}. When it is possible to resolve \hi\ absorption at kpc scales, we can extract information that is similar to what we obtain from \hi\ emission studies on the distribution and kinematics of cold gas at galactic scales. Yet, there have only been a few studies so far where a galaxy has been imaged in \hi\ absorption at kpc scales; for example, the intervening spiral galaxies towards 3C\,196 at \textit{z} \apx 0.4, PKS\,1229$-$021 at \textit{z} \apx 0.395 \citep{Kanekar04} and the spiral galaxy UGC 00439 at $z \sim 0.02$ of a quasar-galaxy pair \citep{Dutta16}. A great deal can also be inferred even when the absorption itself is unresolved, but knowledge of the background radio structure at a higher spatial resolution is available, which can be used to constrain the nature of the absorbing gas via modelling \eg{Briggs99, Briggs01, Murthy19}.

With the commencement of large, deep, blind \hi\ surveys planned with the Square Kilometre Array (SKA) and its pathfinder and precursor facilities, which are capable of simultaneously covering a large redshift range, there is the potential to expand such studies to large numbers and to gain particularly valuable insights at higher redshifts.

Here, we present an example of such a study of resolved \hi\ absorption towards the radio galaxy \target. As we detail below, \target\ has a highly asymmetric radio morphology with most of the flux density arising from the southern lobe. The \hi\ absorption was detected towards \target\ by \citet{Mirabel89} using the Arecibo telescope. Due to the brightness asymmetry in the radio continuum, the authors proposed that the absorption is likely to arise against the southern lobe. Since the large size of the lobe could provide an extended background, this system is a good candidate for carrying out a resolved \hi\ study in absorption.

\begin{figure*}
    \includegraphics[width=\linewidth]{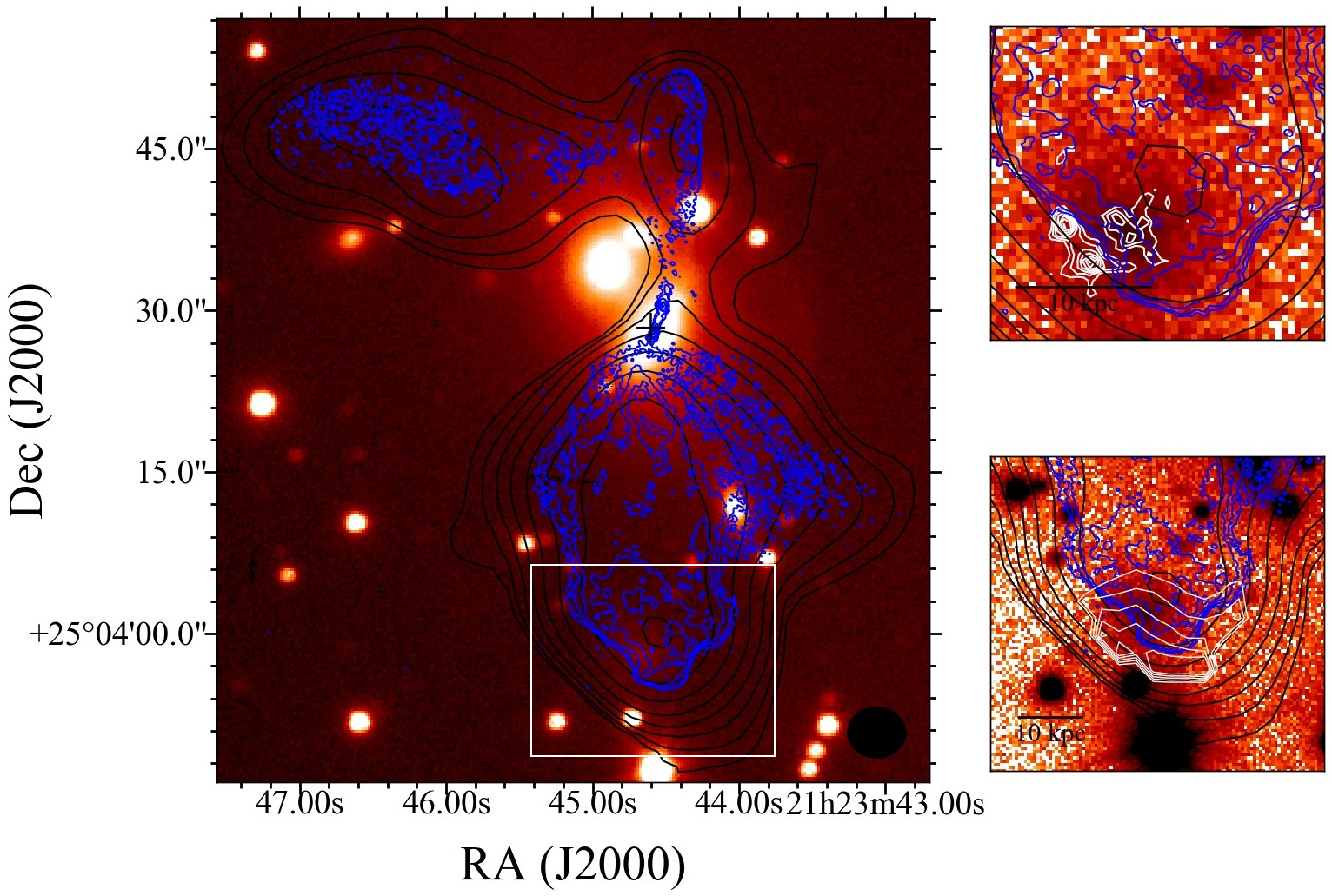}
    \caption{\textbf{Left panel:} Optical image of \target\ environment from GTC is shown in the background. Black contours show the radio continuum at 1.12 GHz as observed with the VLA.  Contours start from 4$\sigma$ (13 mJy beam\p{1}) and increase by a factor of 2. Our VLA radio image has a beam of 5.39$^{\prime\prime} \times$ 4.72$^{\prime\prime}$, shown in the bottom right corner. Blue contours show the 3.6 cm map by \citet{Black92}. Contour levels start from 30$\rm\mu$Jy beam\p{1} (3$\sigma$) and increase by a factor of two. The region in the white box is blown up in the two images on the right panel. The radio core of \target\ as identified by \citet{vanBreugel83} is marked with a black cross. \textbf{Right panel (bottom):} Blow-up of the image in the left panel with the \hi\ column density (N$_{\rm HI}$) contours shown in white. The N$_{\rm HI}$ contours start from 0.7 $\times 10^{18}$(\tspin/f) cm\p{2} and increase in steps of 0.3 $\times 10^{18}$(\tspin/f) cm\p{2}. The image clearly shows that the \hi\ absorber overlaps with an optical galaxy. \textbf{Right panel (top):} \halpha\ contours (from GTC) overlaid on the optical image. The \halpha\ contours levels are: (2.7, 3.2, 3.7, 4.3, 4.9, 5.4, 6 and 6.4) $\times$ 10\p{18} erg s\p{1} cm\p{2}. The 3.6cm radio contours (same as that in the left panel) are shown in blue.}
    \label{fig:optical_radio_halpha}
\end{figure*}

The peculiar radio morphology of \target, as seen in Fig. \ref{fig:optical_radio_halpha}, has already been discussed in various studies \eg{vanBreugel83, Parma91, Black92, Leahy97}. It has a projected angular size of 58$''$, corresponding to a linear size of 110 kpc. It has a weak radio core at the base of a highly collimated jet expanding to the north. The jet is initially slightly curved and has a gap in the middle. There is no similar jet-like feature to the south. Instead, the southern lobe has a large opening angle (80$^\circ$) right at the beginning and contributes to \apx 85\pc\ of the radio emission from this source at 1.4 GHz. 
The southern lobe does not exhibit a relaxed morphology but, instead, it protrudes at the end of the lobe as if it has been pinched. It also contains a hotspot and various complex and fine structures \citep[see e.g. Fig. 19 in][]{Leahy97}. Thus, in the \citet[][FR]{Fanaroff74} classification scheme, \target\ is a hybrid-morphology source exhibiting an FR I radio jet and an FR II radio lobe. In addition, the southern lobe has a faint wing on the western side, which appears to lie at the same angle from the core as the outer emission in the northern lobe. This emission gives \target\ an X-shaped appearance \eg{Lal07, Gillone16}.

The host galaxy of the \target\ is a part of an interacting pair, enclosed in a common envelope \citep[see Fig. \ref{fig:optical_radio_halpha};][]{Matthews64, vanBreugel83, Baum88, Smith89, Black92}. It has been found to have young stellar population with ages  $0.03 < \rm t_{\rm YSP} < 0.1$ Gyr \citep{Tadhunter11}, which is very likely formed due to the ongoing interaction. \citet{Miller09} have carried out X-ray observations with \textit{Chandra} and find diffuse X-ray emission in the soft-band, which curves along the east side of the southern lobe in the 0.5 - 2 keV smoothed image.

In order to localise and, if possible, resolve the \hi\ absorber, we observed \target\ with NSF's Karl G. Jansky Very Large Array (VLA) in the B configuration. Furthermore, to better characterise the absorber, we also obtained deep, narrow, and medium band optical continuum and \halpha\ images of the field with the Gran Telescopio CANARIAS (GTC). Interestingly, we find that the absorption is due to an intervening disc galaxy with a redshift close to that of \target. This allows us to study  the properties of the galaxy in detail and explore the possibility of an interaction between the radio lobe of \target\ and the galaxy, as well as its effect on the observed morphology of the radio lobe. 

We describe our radio and optical observations in Section 2, present our results in Section 3, and discuss the possible origin of the \hi\ absorption and its implications in Section 4. Finally, we summarise our findings in Section 5. We have assumed a flat universe with H$_0$ = 67.3 km s$^{-1}$Mpc$^{-1}$, $\Omega_{\Lambda}$ = 0.685, and $\Omega_M$ = 0.315 \citep{Planck14} for all our calculations. \target\ is at \textbf{$z = 0.1016 \pm 0.0001$} \citep[an optical systemic velocity of 30458.9 \kmps;][]{Schmidt65, Hewitt91} where 1$''$ corresponds to 1.943 kpc.

\section{Observations and data reduction}\label{data}
\label{methods}

\subsection{VLA observations}

Our VLA observations were carried out in July 2002 with antennas in B configuration (project id: AM0730). We observed \target\ for 4.5 hours in total. Because the observations were performed before the upgrade to the VLA wideband WIDAR correlator, only a single Stokes parameter (RR) was used in order to cover a sufficiently wide bandwidth of 6.2 MHz subdivided into 127 channels. The observations consisted of interleaved scans on the flux and bandpass calibrator (3C\,48), the phase calibrator (B2 2113+29), and the target.

The data reduction was done in `classic' AIPS (Astronomical Image Processing Software). We first flagged the bad baselines and bad data. Then determined the antenna-dependent gain and bandpass solutions using the data on calibrators. We then iteratively improved the gain solutions via self-calibration. Initially, we carried out a few cycles of imaging and phase-only self-calibration. Then carried out a round of amplitude and phase self-calibration and imaging. We then subtracted the continuum model from the calibrated visibilities and flagged the residual UV data affected by radio frequency interference. We fit a second-order polynomial to the line-free channels of each visibility spectrum. Finally, we imaged this UV data to get the spectral cube. 

We made the continuum map using robust weighting of -1, averaging all the line-free channels together. It
has a restoring beam of 5.39$''$ $\times$ 4.72$''$ with a position angle of -89.27$^\circ$ and has an RMS noise of \apx 2.6 mJy beam\p{1}. We made the spectral cube with the same weighting and the same restoring beam as the continuum map. It has an RMS noise of \apx1.2 mJy beam\p{1} channel\p{1} for a channel width of 12.5 \kmps, without any spectral smoothing. The \hi\ moment maps (shown in Fig. \ref{fig:optical_radio_halpha} and Fig. \ref{fig:velocity_field}) were produced from this spectral cube by adding the channels across which the absorption was found. 

The peak flux density of the target is 1.3 $\pm$ 0.1~Jy~beam\p{1}. The integrated flux density is  14.2 $\pm$ 0.7 ~Jy. The southern lobe contains \apx 85\pc of the total flux density (\apx 13 Jy). The uncertainty on the flux density scale at the observed frequency is assumed to be 5\pc \citep{Perley2017}.

More recent VLA observations of \target\ from 2017 (project id: 17B-016) are available in the archive. A part of these observations was carried out in spectral line mode with a 16 MHz band, centred at the redshifted \hii\ frequency, subdivided into 1024 channels. Thus, these data have a significantly better spectral resolution that would enable us to study the absorption profile in greater detail. Unfortunately, these observations are severely affected by RFI and the data are unusable.

\begin{figure*}
    \includegraphics[width=\linewidth]{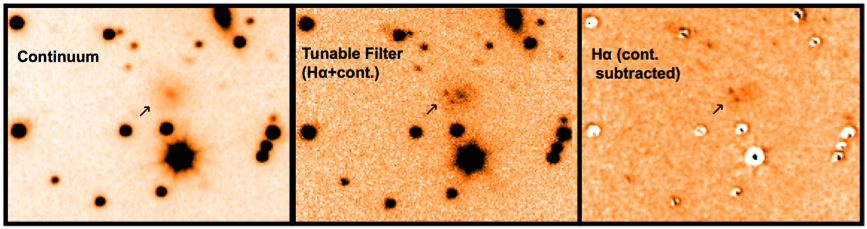}
    \caption{GTC imaging of the galaxy at the location of the \hi\ absorption. Shown: the r-band continuum image (left), the narrow-band image with both H$\alpha$ and continuum included (middle), and the continuum-subtracted H$\alpha$ image (right). The galaxy is indicated by the black arrow in each panel. The details on these images are given in Sect. \ref{results}. }
    \label{fig:GTCcompanion}
\end{figure*}

\subsection{Optical and \halpha\ data}\label{optical_data}

Optical narrow and medium band images of 3C\,433 were taken with the GTC on 11 Sept 2017 (project GTC48-17B), using the Optical System for Imaging and low-Intermediate-Resolution Integrated Spectroscopy (OSIRIS). The H$\alpha$ emission line was captured using the tunable narrow-band filter f723/45, centred at a wavelength of 7234 \AA\ with a full width at half the maximum intensity (FWHM) of 20 \AA. The continuum emission was observed with the medium-band order sorter filter f666/36, which has a central wavelength of 6668 \AA\ and an FWHM of 355 \AA. This corresponds to r-band wavelengths and is free of emission lines. The total on-source exposure time was 1\,h in the narrow-band filter, divided into 15 dithered exposures of 250\,sec. For the medium-band filter, this was 25\,min, with 15 dithered exposures of 100\,sec. The observations were performed in queue mode under seeing conditions of roughly 1$''$. 

We reduced the data using the Image Reduction and Analysis Facility \citep[\textsc{IRAF};][]{tody86}. After a standard bias subtraction and flat-fielding, we co-added the images in each filter while removing cosmic ray. Because the wavelength tuning is not uniform for the OSIRIS narrow-band filters, the effective field of view is significantly smaller than the unvignetted 7.8$^{\prime}$ coverage of the CCD, and background gradients are introduced in the narrow-band imaging. We removed these background gradients as best as possible by fitting a 2-D polynomial to the background emission and subtracting this from the narrow-band image. We used the same technique to produce the medium-band image of the continuum. We then subtracted the medium-band continuum image from the narrow-band image to obtain an image with only H$\alpha$ emission. We did this by scaling down the raw counts of the continuum image by a factor 5.75 (derived empirically) and then subtracted that from the narrow-band image. In the resulting H$\alpha$ image, any remaining background gradient was removed by fitting the background separately across RA and Dec while manually excluding emission from galaxies, stars and artefacts, and then subtracting a 2D average of these RA and Dec fits from the H$\alpha$ image. Due to the large uncertainty in absolute astrometry of the GTC data, we shifted the final GTC images by 2$^{\prime\prime}$ to align with the existing Hubble Space Telescope (HST) imaging. We would like to note here that the artefacts associated with the point or highly nucleated sources seen in the \halpha\ images are unavoidable. Spatial shifts of a small fraction of a pixel or small differences in seeing size between the continuum and narrowband images (or both) will always produce such artefacts. 

We do not have an absolute calibration for our GTC images since we did not observe a standard star, as that would have resulted in a large increase in the overhead time for these queue-mode observations. Fortunately, a part of the field covered by our observation has been observed by the Sloan Digital Sky Survey (SDSS) and, hence, the SDSS stars could be used for calibration. To this end, we first extracted the point sources in our image using SE\textsc{xtractor} \citep{Bertin96}. The software generated a catalogue of sources from which we selected those sources with CLASS\_STAR $\gt$ 0.85 and FLAG $\lt$ 4. This was done to ensure that we selected only the point sources which are not saturated. We extracted the instrumental magnitudes for these stars from SE\textsc{xtractor} via the MAG\_AUTO parameter. We cross-matched the stars thus obtained with the stars in the SDSS Data Release 12 r-band catalogue \citep{Alam15}. Thence, we obtained the zero-point magnitude by taking the difference between the SDSS magnitude and the magnitude obtained from SE\textsc{xtractor}. The final zero-point magnitude is the mean of the zero-point magnitudes thus obtained. We obtained a final zero-point magnitude of 26.71 \plm\ 0.49 for our continuum image. The GTC images, corrected for the astrometric offset using the HST imaging as mentioned above, align with the SDSS stars. 

As we go on to explain in the  sections to follow, a faint galaxy located close to the edge of the southern radio lobe of \target\ is of particular interest for this work. Figure \ref{fig:GTCcompanion} shows the medium-band, narrow-band, and H$\alpha$ GTC images of this galaxy. We used \textsc{GALFIT} \citep{Peng02, Peng10} to characterise this galaxy. We used a cutout of the field (123 $\times$ 134 pixels), including the galaxy, as the input to \textsc{GALFIT}. We fit for the centre of the galaxy, integrated r-band magnitude, effective radius (r$_e$), axial ratio, and the sky background. The initial guess for the centre of the galaxy, effective radius, and the axial ratio were determined visually. The initial guess for the integrated r-band magnitude was input from the Panoramic Survey Telescope and Rapid Response System (Pan-STARRS) r-band catalogue \citep{Chambers16, Flewelling16} in which the galaxy is detected.  We fine-tuned the initial parameters over a few iterations and found that the final fit values agreed within the error bars. This galaxy has also been detected in the Pan-STARRS g-band image. We extracted the g-band integrated magnitude of the galaxy following the same procedure as mentioned for our continuum image using SE\textsc{xtractor} and \textsc{GALFIT} on the Pan-STARRS image. We have summarised the fit parameters in Table~\ref{table:galfit}.

Then we performed aperture photometry in continuum using Photoutils of Astropy package under Python software \citep{Bradley19}. We used circular apertures of radii varying from 0.1r$_e$ to r$_e$. We did not extend beyond the unit effective radius for the presentation of the surface brightness profile due to the presence of a bright source to the south of the galaxy whose contribution would become significant beyond that distance. The choice of circular aperture was motivated by the high axial ratio of the galaxy (see Sect. \ref{results}) as obtained from \textsc{GALFIT}. 
We discuss the parameters obtained in Sect. \ref{sec:counterpart}.

\begin{table}[]
\begin{tabular}{cccc}
\hline
Filtre & Magnitude    & R$_{eff}$ ($''$)     & Axial ratio \\
(1)& (2) & (3) & (4)\\
\hline
r      & 16.61$\pm$0.15 & 3.3$\pm$0.4 & 0.88$\pm$0.02  \\
g      & 17.22$\pm$0.08  & 2.6$\pm$0.3 & 0.93$\pm$0.07 \\
\hline
\end{tabular}
\caption{Summary of the GALFIT parameters. The columns are: (1) Filtre; (2) Integrated apparent magnitude; (3) Effective~radius~($''$); (4) Axial ratio}
\label{table:galfit}
\end{table}

\begin{figure}
    \includegraphics[width=\linewidth]{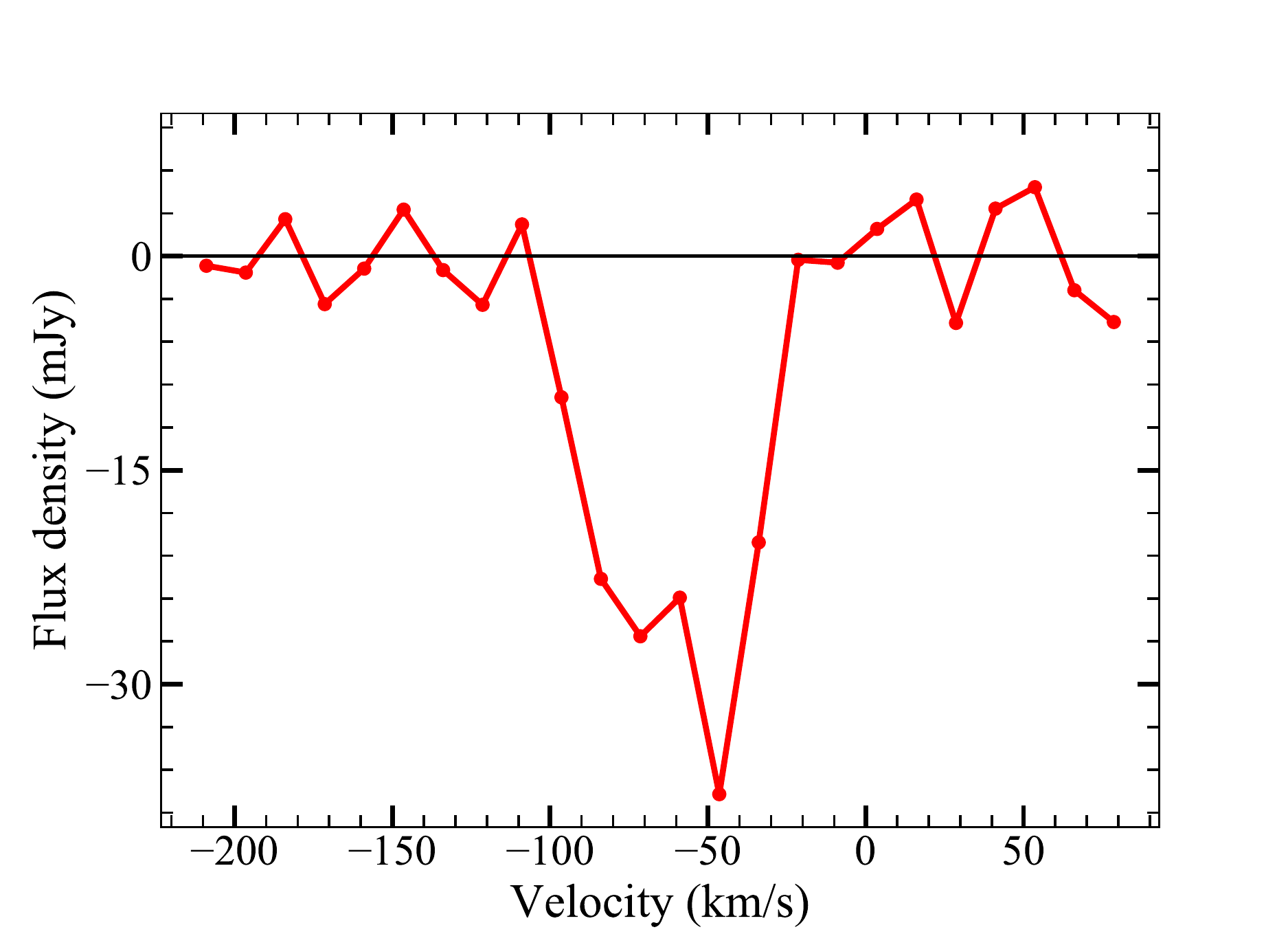}
    \caption{Integrated \hi\ absorption spectrum from the VLA. The spectrum has been shifted to the restframe of \target\ i.e. $z = 0.1016 \pm 0.0001$. We see that the peak absorption is only \apx 50 \kmps blueshifted with respect to the systemic velocity of \target.  The VLA spectral cube has an RMS noise of 1.2 mJy beam$^{-1}$ and a velocity resolution of 12.5 \kmps.}
    \label{fig:int_spc}
\end{figure}

\begin{figure}
    \includegraphics[width=\linewidth]{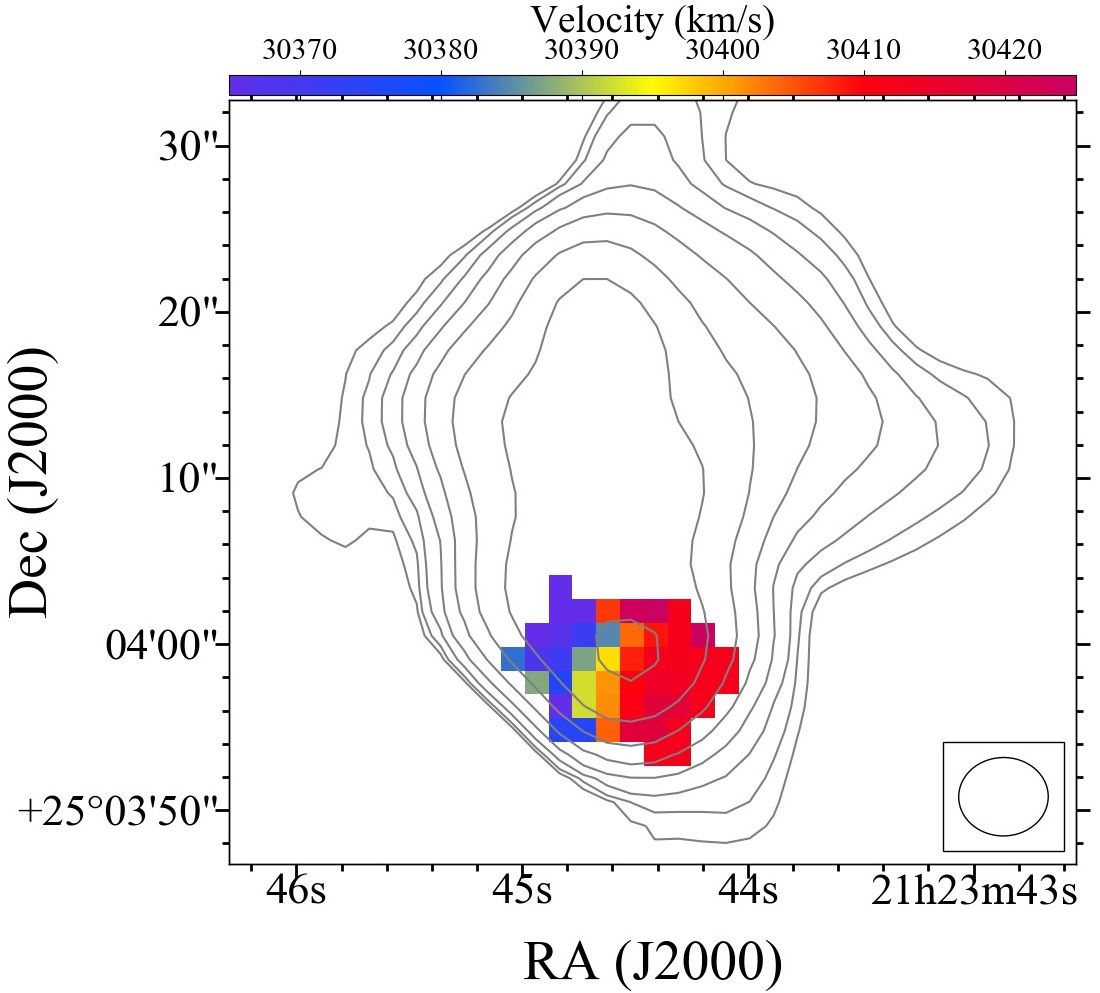}
    \caption{\hi\ absorption velocity field with the contours of the radio continuum of \target\ (from the VLA observations) overlaid. Contour levels and the beam (shown in the bottom right corner) are as in Fig. \ref{fig:optical_radio_halpha}.}
    \label{fig:velocity_field}
\end{figure}

\begin{figure}
    \includegraphics[width=\linewidth]{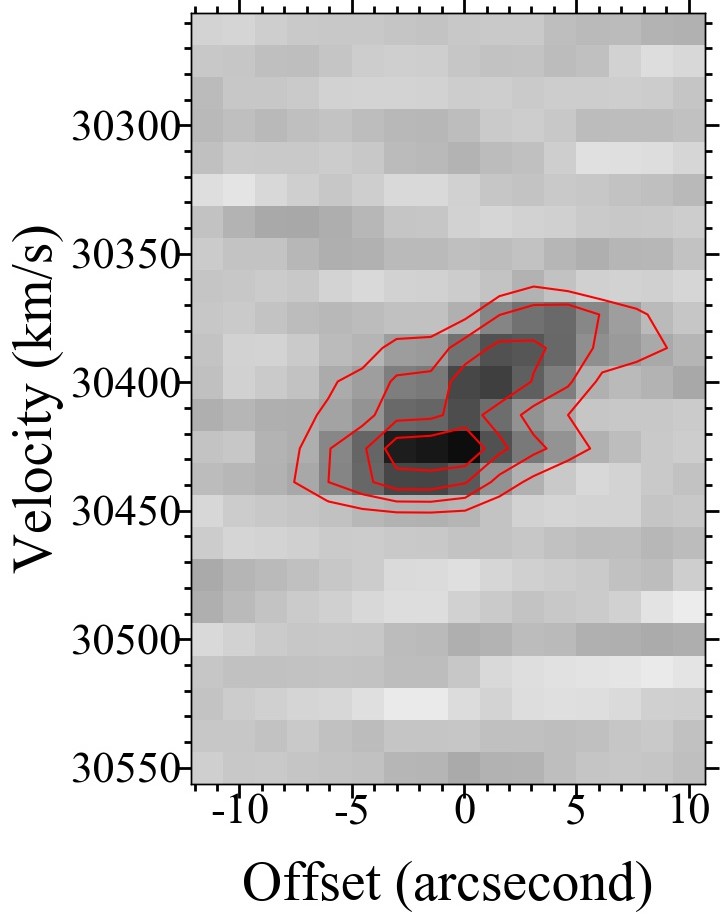}
    \caption{\hi\ position-velocity diagram of the absorber galaxy extracted along the major axis of the \hi\ absorption (i.e. \apx 63$^\circ$, visually determined). Contour levels are: 3.5, 7.3, 11.7, 16.0, and 20.4 mJy beam\p{1}. The systemic velocity of \target\ is 30458.9 \kmps.}
    \label{fig:pvd}
\end{figure}

\section{Results}\label{results}

The radio and the optical data both show interesting and new features. We describe them below, including the optical properties of both the host galaxy of 3C~433 and its environment.  We then focus only on the \hi\ absorber in the discussion in Sect. \ref{discussion}.

\begin{figure*}
    \includegraphics[width=\linewidth]{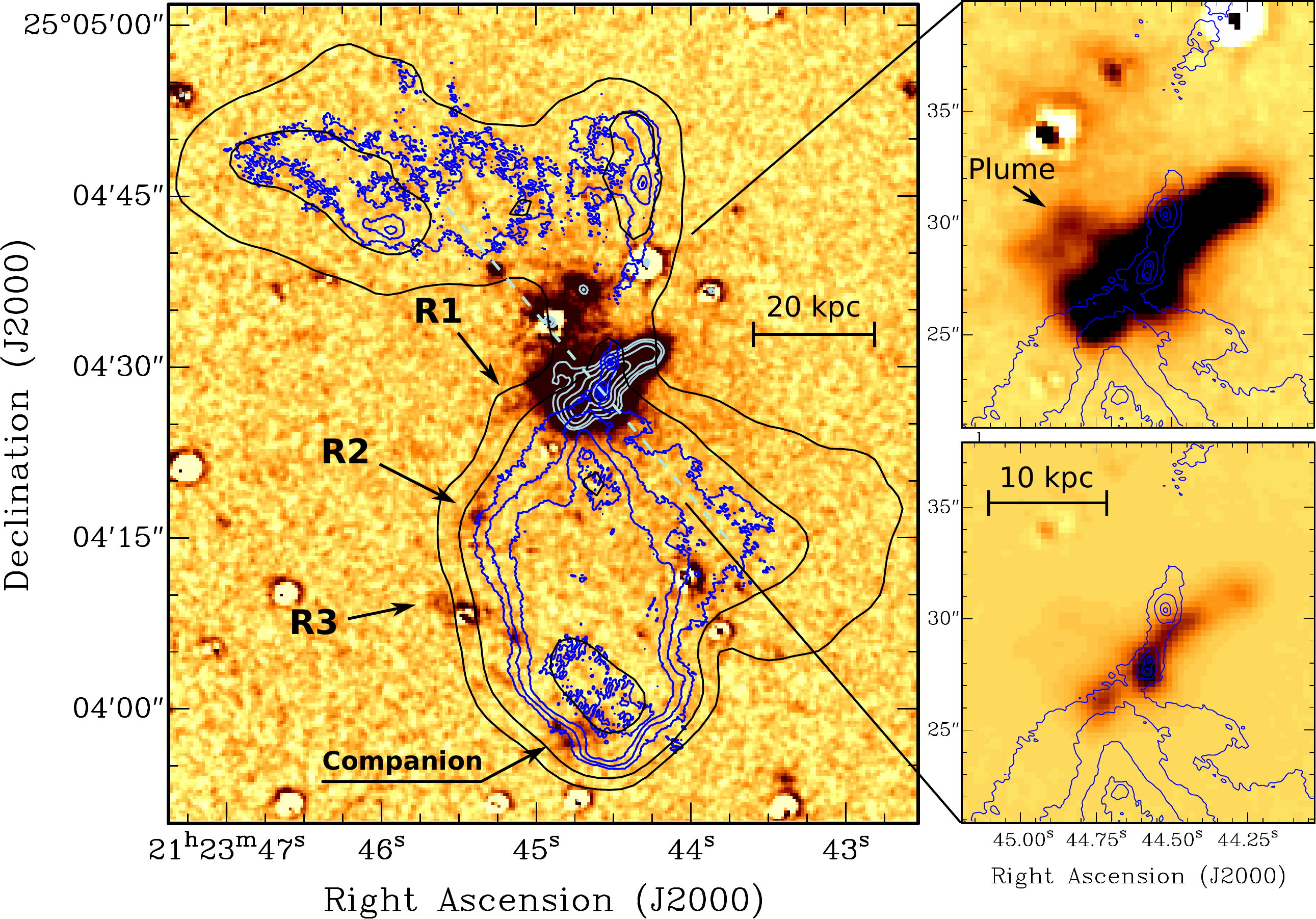}
    \caption{\halpha\ image of 3C433 and its environment. \textbf{Left:} Low surface-brightness emission in the circum-galactic environment of \target, showing the companion galaxy to the south and several H$\alpha$ emitting regions. The white dots are mostly the residuals of saturated foreground stars that could not be accurately subtracted. The black and dark blue contours show the low- and high-resolution VLA radio continuum images previously shown in Fig. \ref{fig:optical_radio_halpha}. The contour levels are at 0.2, 0.5, 1.0, 1.5, 2.0 mJy beam$^{-1}$ for the high-resolution data (blue) and 15, 60, 700 mJy beam$^{-1}$ for the low-resolution data (black). Light blue contours highlight the H$\alpha$ emission shown in more detail on the right-hand side. The contour levels are at 4.3, 6.4, 12.8, 25.7, 47 percent of the peak intensity. Light-blue dashed line visualizes the direction perpendicular to the \halpha\ structure in the center. \textbf{Right:} \halpha\ emission in the central radio galaxy, shown at two different contrasts.}
    \label{fig:Halpha}
\end{figure*}

\subsection{Radio continuum and the HI absorption}\label{results_radio}

Our VLA continuum map of \target\ is shown in Fig. \ref{fig:optical_radio_halpha}, left panel in black contours. Due to low spatial resolution, the fine structures and the hot-spot in the southern lobe mentioned in Sect. \ref{introduction} are smoothed out in our continuum map. However, the X-shaped morphology and the brightness difference between the northern and southern lobes are evident.

As expected, we detected the \hi\ absorption and, interestingly, we find that the absorption is confined to the southern part of the southern lobe. The location of the \hi\ absorption is indicated by the white contours in Fig. \ref{fig:optical_radio_halpha} (bottom-right panel). As mentioned in Sect. \ref{introduction}, \citet{Mirabel89} had suggested that the \hi\ absorption may arise towards the southern radio lobe since it contains most of the flux. Our observations clearly show that the \hi\ is confined (in projection) to a small part of the lobe. This is essential for the interpretation of the origin of the absorption, as we discuss in Sect. \ref{discussion}. The integrated \hi\ absorption profile extracted from the region detected in \hi\ absorption is shown in Fig. \ref{fig:int_spc}. The absorption is \apx50 \kmps\ blue-shifted from the systemic velocity of \target,  which is in agreement with that reported by \citep{Mirabel89}. The integrated absorption line is \apx 40 mJy deep and has a full width at zero intensity (FWZI) of  \apx 80 \kmps. 

The \hi\ column density distribution is shown in Fig. \ref{fig:optical_radio_halpha}, bottom-right panel. The average \hi\ column density is (1.0 $\pm$ 0.2) $\times$ 10$^{18}$ (\tspin/f) cm\p{2} where \tspin\ is the gas spin temperature and f is the covering factor which, in this case, is unity on the account of the absorber being resolved. The absorber is \apx 60 kpc, in projection, from the radio core of \target\ identified by \citet{vanBreugel83}. The velocity field is shown in Fig. \ref{fig:velocity_field} and Fig. \ref{fig:pvd} shows the position-velocity (PV) diagram along the major axis of the absorber. We find that the velocity gradient is smooth throughout the structure, and is probably related to rotation. 

There is a steep gradient in the brightness of the southern lobe and the flux density decreases  away from the region overlapping with the detected \hi. The average flux density in the region immediately northwards of the absorber is \apx 700 mJy beam\p{1}. Using this and assuming an absorption profile of the same FWHM as the detected integrated absorption (60 \kmps), we derive a 3$\sigma$ \hi\ column density sensitivity of 7.5 $\times$ 10\pp{17} (\tspin/f) cm\p{2} towards regions northwards of the absorber. Similarly, the average flux density to the south of the absorber is \apx 50 mJy beam\p{1}. That gives us a 3$\sigma$ sensitivity of 9 $\times$ 10\pp{18} (\tspin/f) cm\p{2} in the regions south to the absorber. Thus, if there is more of \hi\ in the region northwards that is covered by the radio continuum, we would have been able to detect it but not towards the southern regions, where the continuum emission drops rapidly.

\subsection{Optical and H$\alpha$ counterparts of the \hi\ absorber}
\label{sec:counterpart}

As can be seen from Fig. \ref{fig:optical_radio_halpha}, a faint galaxy is present, in projection, at the location of the \hi\ absorption and, hence, it is of interest as a possible candidate giving rise to the absorption. Because of this, we characterised this galaxy further using the methods described in Sect. \ref{data}. The galaxy is shown clearly in Fig. \ref{fig:GTCcompanion} and has an effective radius, r$_{e} = 3.3^{\prime\prime}$, that is, 6.3 kpc and an axial ratio of 0.88. Fig. \ref{fig:surface_brightness} shows the surface brightness profile of this galaxy up to one effective radius. We find that an exponential disc fits the surface brightness profile very well. 

We measured an r-band integrated apparent magnitude of m$_r = 16.61$. As mentioned in Sect. \ref{data}, this galaxy has also been detected by Pan-STARRS in g-band with an integrated apparent magnitude, m$_g = 17.22$. Thus, the K-corrected \citep{Chilingarian10,Chilingarian12} g$-$r colour of the galaxy is 0.46. Using the correlation between the mass-to-light ratio and the K-corrected g$-$r colour given by \citet{Bell03}, we obtain a stellar mass \Mstar\ \apx 6.7 $\times$ 10\pp{10} \Msun\ for the galaxy.

Our GTC observations reveal an interesting morphology for the \halpha\ emitting gas in this galaxy as shown in Fig. \ref{fig:optical_radio_halpha} top-right panel. We detected \halpha\ emission from the central stellar region of the galaxy and also along the eastern edge. It is worth noting that this outer \halpha\ blob coincides with the outer boundary of the radio lobe (see Fig. \ref{fig:optical_radio_halpha}). We  come back to this aspect in Sect. \ref{discussion}. One possibility is the presence of an interaction between the radio lobe and the galaxy, and that this component of \halpha\ emission is due to shock excitation of the gas resulting from this.

The total \halpha\ flux from the galaxy is 5 $\times$ 10\p{16} erg cm\p{2} s\p{1} while the emission from only  the central region is \apx 2 $\times$ 10\p{16} erg cm\p{2} s\p{1}. With these \halpha\ fluxes, we can estimate the star-formation rate (SFR) using the Kennicutt-Schmidt law \citep{Kennicutt98}. Assuming that all the \halpha\ emission detected from the galaxy arises from star formation activity, we obtain a star-formation rate (SFR) of 0.11 \Msun\ yr\p{1}.  If only the \halpha\ emission arising from the centre of the galaxy is due to star formation (while the eastern \halpha\ component is due to shock excitation), we obtain a lower limit to the SFR of 0.035 \Msun\ yr\p{1}. The SFR is lower even if we are to infer it from other tracers such as the infrared (IR) luminosity: following \citet{Kewley02}, we use the relation SFR~(IR)~\apx~2.7~ $\times$~SFR~(\halpha)$^{1.3}$ and find that SFR(IR) ranges from 0.15 \Msun\ yr\p{1} to 0.034 \Msun\ yr\p{1} for the two cases mentioned above. 

We note here that the \hi\ centre of the galaxy is offset from the optical centre (Fig. \ref{fig:optical_radio_halpha}, bottom-right panel). The left panel of the same figure shows the location of the optical galaxy with respect to the radio lobe (at higher spatial resolution than our VLA map; blue contours). We find that the radio continuum does not intercept the galaxy in its entirety since the latter is located at the boundary of the former. Therefore, it is likely that a part of the \hi\ disc could be missing due to the absence of background radio continuum in that region. Thus, the mismatch between the \hi\ and optical centres could be due to the non-detection of a part of the \hi\ disc.

\subsection{H$\alpha$ in the host galaxy of \target}

\target\ is associated with one of the galaxies in an interacting pair surrounded by a common envelope \citep{Matthews64}. Earlier optical broad-band, narrow-band \halpha\ and spectroscopic studies of the system have shown clear \halpha\ filaments, a young stellar population, and signs of recent disturbance in the form of distorted dust lanes \citep[e.g.][]{Baum88, deKoff96, deKoff00, Holt07, Tadhunter11}.  

Our optical image of \target\ and its environment is shown in Fig. \ref{fig:optical_radio_halpha}. It is of much better sensitivity and has a superior angular resolution compared to the Pan-STARRS r-band image. We clearly see the optical counterpart of \target\ along with its companion galaxy. Additionally, we also clearly detect shells around this interacting pair, very likely caused due to their interaction.

Figure \ref{fig:Halpha} shows the H$\alpha$ emission-line image of radio galaxy \target\ and its circumgalactic environment. The H$\alpha$ emission at the location of the host galaxy of \target\ shows an elongated structure, previously seen by \citet{Baum88}. The total extent of the H$\alpha$ structure is $\sim$10$^{\prime\prime}$ ($\sim$19\,kpc). The brightest patch of H$\alpha$ emission corresponds to the peak in the continuum image, hence we argue that this is most likely the core of \target.

\begin{figure}
    \includegraphics[width=\linewidth]{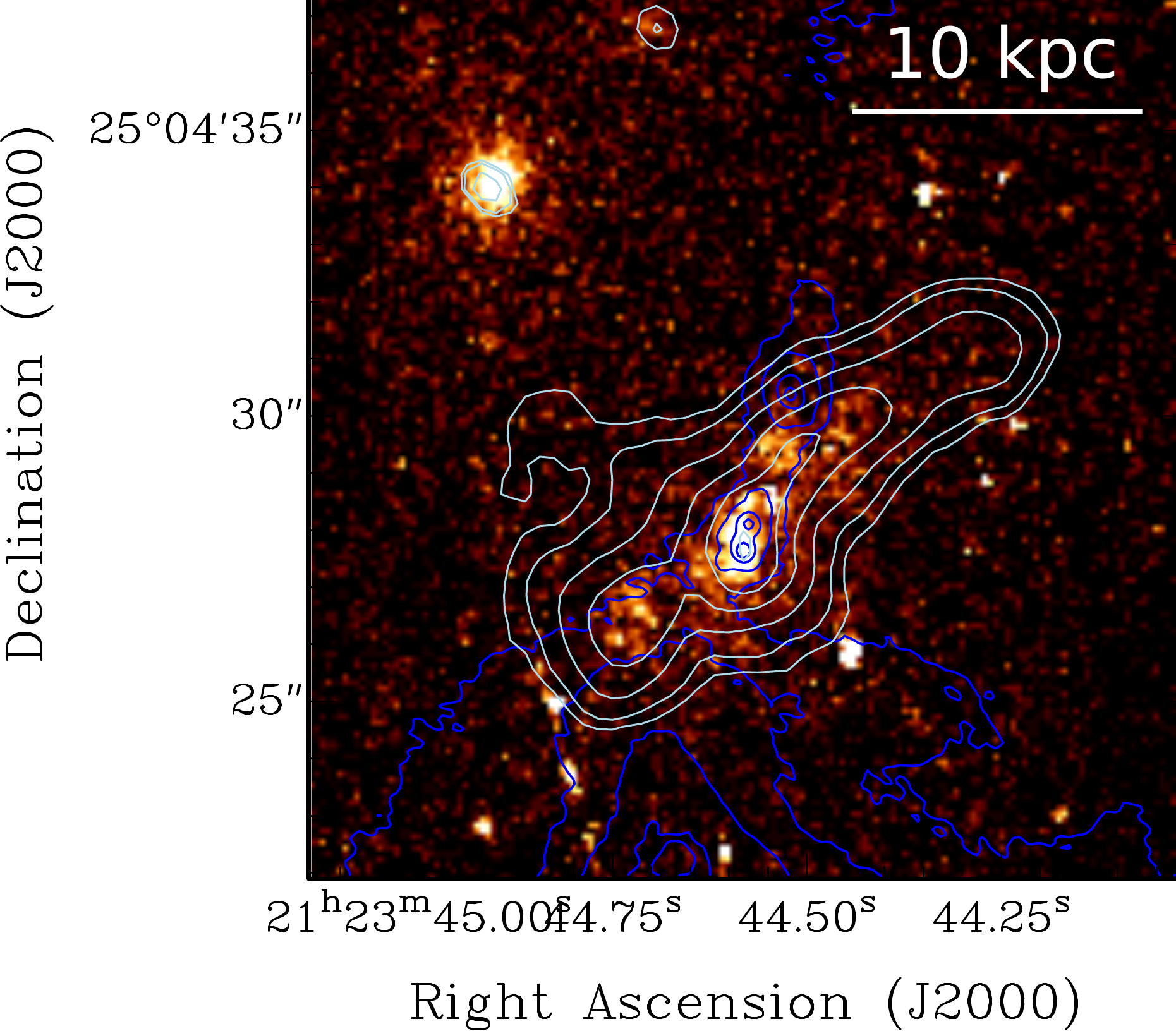}
    \caption{HST image of [OIII] detected with filter FR533N18 centered at 5515.7$\AA$ ($\nu_{\rm rest}$\,=\,5007$\AA$).  Overlaid are the contours of our H$\alpha$ image (light blue) and VLA radio continuum (dark blue) from Fig. \ref{fig:Halpha}. The H$\alpha$ emission follows the [OIII] emission, which, in turn, traces the outline of the inner radio jet.}
    \label{fig:overlayHST}
\end{figure}

A plume of H$\alpha$ emission is seen to stretch roughly 4$^{\prime\prime}$ ($\sim$8\,kpc) north-east of the centre of the radio galaxy (see image in Fig. \ref{fig:Halpha}, top-right panel). This plume extends almost perpendicular to the radio axis. It could represent gas that is either being driven out of or being accreted onto the centre of the galaxy. Further to the north, at 9$^{\prime\prime}$ ($\sim$17\,kpc) distance from the radio galaxy, another region of enhanced emission appears in the H$\alpha$ image. This region aligns with what appears to be a companion galaxy in the continuum image. It is not clear whether this emission is from the H$\alpha$-emitting gas in this companion galaxy, or residual emission left over from the continuum subtraction.

Fig. \ref{fig:overlayHST} shows the H$\alpha$ emission in the centre of \target\ overlaid onto a Hubble Space Telescope narrow-band image of [OIII] emission. The H$\alpha$ emission shows a distribution that is very similar to the [OIII]-emitting gas in the inner ~12 kpc. In turn, the [OIII] emission traces the inner radio jet. It is therefore likely that both the [OIII] and a large fraction of the H$\alpha$ emission in the inner 12 kpc trace gas that is under the influence of the propagating radio jets. It is plausible that the radio jets have shock-ionized the gas, or even triggered star formation.

In addition to the co-alignment of H$\alpha$ and [OIII] in the inner 12 kpc, the H$\alpha$ stretches further out to the north-west than the [OIII], in what was classified as a curving filament by \citet{Baum88}. The overall distribution of the H$\alpha$ emission resembles a gaseous disc with an extent of $\sim$19 kpc, but this would have to be verified by observing the kinematics. If confirmed, this structure would be misaligned with the main radio axis by $\sim$40$^{\circ}$. However, it is interesting to note that the outer faint wings of radio emission that give the \target\ its X-shape, lie at an angle of \apx 90$^{\circ}$ with the optical structure and, thus, parallel to its minor axis, possibly corresponding to the rotation axis (see light-blue dashed line in Fig. \ref{fig:Halpha}). This would suggest that the central black hole and its accretion disk underwent a significant precession recently, possibly during the current episode of radio-AGN activity, whereby the jet direction changed by $\sim$50$^{\circ}$ in the plane of the sky. It is quite possible that such a precession may have been triggered by the ongoing galaxy interactions, and is the mechanism \citep[of the many mechanisms proposed to explain the X-shape in radio galaxies;][]{Cotton20, Hardcastle19} responsible for the observed X-shape of \target.

\begin{figure}
    \includegraphics[width=\linewidth]{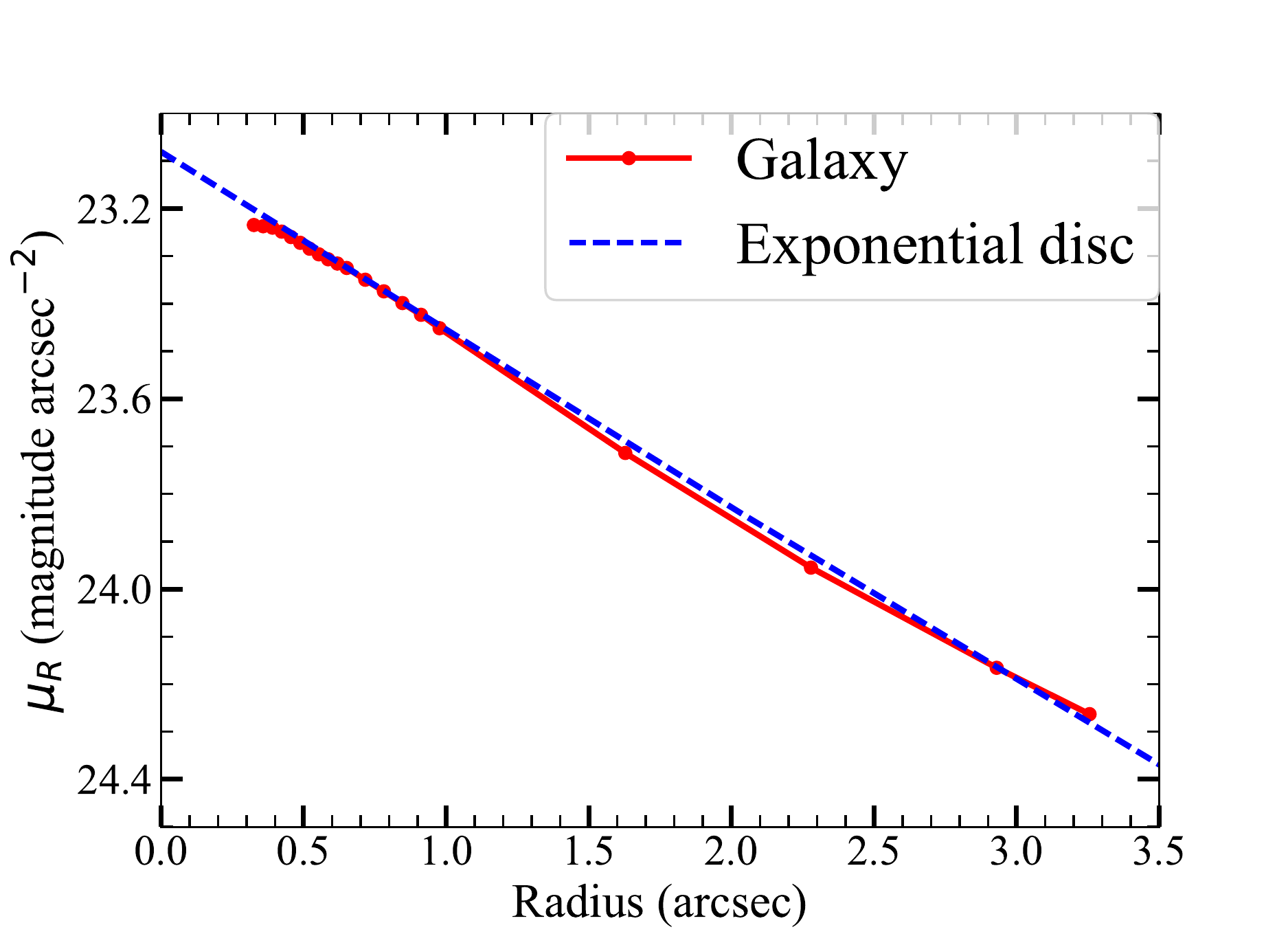}
    \caption{Surface brightness profile of the absorber galaxy. For representation, we limit this to one effective radius. See Sect. \ref{results} for details.}
    \label{fig:surface_brightness}
\end{figure}

\subsection{The circumgalactic environment of \target}

The \halpha\ image of Fig. \ref{fig:Halpha} shows several other interesting features. At least three blobs of H$\alpha$ emission are seen at distances of roughly 8, 23, 15, and 23 arcsec (16, 29, 45 kpc) from the centre of the radio galaxy. These blobs are marked with a black arrow in Fig. \ref{fig:Halpha}. Regions R1 and R2 only appear in the tunable filter image (H$\alpha$+continuum) and not in the continuum image, while the southernmost region (R3) has a higher contrast in the tunable filter compared to the continuum image. These three regions of enhanced H$\alpha$, therefore, most likely represent real emission-line regions. There may be other regions of H$\alpha$ emission, for example, $\sim$7$^{\prime\prime}$ south-west of R3, but due to stronger underlying continuum at those locations, we cannot confirm this.

The northernmost region (R1) appears to be a weak emission-line feature stretching from the radio host galaxy. Interestingly, the three emission-line regions lie along an arc that follows the outer edge of the southern radio lobe. This could be tidal debris from material that is being redistributed across the circum-galactic environment of \target. We hypothesise that the alignment with the outer edge of the radio source may occur because this circum-galactic material is being shocked and ionized by the propagating radio source, similar to what has been seen in Coma A \citep{Morganti02} and the Beetle Galaxy \citep{Villar17}. The H$\alpha$ regions are also aligned in the direction of the \hi\ companion, although a direct connection between this circumgalactic gas and the \hi\ companion is not apparent from our data.

\section{Discussion}\label{discussion}

Our spatially resolved \hi\ absorption observations of \target\ have shown that the absorption arises against the southern radio lobe, about 60 kpc away from the radio core. We find that the central velocity of the \hi\ absorption is only \apx 50 \kmps\ blueshifted from the systemic velocity of \target\ (see Fig. \ref{fig:int_spc}), confirming that it belongs to the same environment. This combination of having the absorption  spatially resolved and the absorber belonging to the same environment, and yet being located so far away from the central region of the background AGN, makes this system a rarity. 

\subsection{The \hi\ absorber: a disc galaxy}

What we observe in \target\ is different from the results commonly obtained from  \hi\ absorption studies of the gas associated with the radio AGN host galaxies. Those studies have mostly found that the \hi\ absorption arises from the central (kpc) regions close to the core of the radio AGN \citep[see][for an overview]{Morganti18}. 
 
Instead, in this case, our optical image shows the presence of a galaxy coincident with the location of \hi\ absorption.  We propose that the \hi\ detected in absorption belongs to this galaxy located in front of the southern radio lobe.  The peak of the \hi\ absorption is only \apx 50 \kmps\ bluewards of the systemic velocity of \target\ and, hence, very likely also belongs to the same environment. The galaxy has a stellar mass of \apx 6.7 $\times$ 10\pp{10} \Msun. It has a surface brightness profile (see Fig. \ref{fig:surface_brightness}) characteristic of a disc galaxy. 

Assuming a spin temperature of \apx 100 K, a value typical of \hi\ on galactic scales \citep[e.g.][and references therein]{Borthakur10, Borthakur14, Srianand13, Reeves15, Dutta16, Reeves16}, we obtain an average \hi\ column density of \apx 1.0 $\times$ 10$^{20}$ cm\p{2}. For spiral galaxies, the average column density of the \hi\ disc is about 6 $\times$ 10$^{20}$ cm\p{2} \citep{Wang16}, quite higher than our absorber galaxy. Instead, \citet{Serra12} found that in gas-rich early-type galaxies, the typical column densities are around 10\pp{20} cm\p{2}, which is very similar to what we observe. This suggests that our galaxy is a gas-rich early-type galaxy, which is also consistent with the fact that it lies off the star-forming main sequence (see Sect. \ref{sec:comparison}). 

Using this \hi\ column density of the gas, we obtain the mass of the detected \hi\ as \apx 3.4 \tim\ 10\pp{8} M$_\odot$. As mentioned in Sect. \ref{results_radio}, it is likely that a portion of the galaxy has not been intercepted by the radio continuum. Since it is only a small part of the galaxy, we argue that the actual \hi\ mass could at most be \apx 7 $\times$ 10\pp{8} \Msun\ (i.e. twice the estimated mass, assuming that half the disc has not been detected). This gives us an atomic gas fraction (f$\rm_{HI}$)   of 0.008 for the galaxy.

As discussed in Sect. \ref{sec:interaction}, there is a possibility of an interaction between the galaxy and the radio lobe. If this is the case, the spin temperature of \hi\ could be higher. Assuming a \tspin\ of 1000 K for \hi, gives an \hi\ mass of \apx\ 3.4 \tim 10\pp{9} \Msun\ and an f$\rm_{HI}$ of 0.08. There is a possibility that the observed absorption could also arise from a tail of tidal debris  formed, for example, due to the ongoing merger event. We exclude this scenario since we do not see the presence of such prominent and large-scale structures in our optical image.

\subsection{Comparison with galaxies observed in \hi\ emission in the local Universe and beyond}
\label{sec:comparison}

As mentioned earlier, studies of spatially resolved \hi\ absorption have been carried out, beyond the local universe, for only two intervening absorbers where the absorber galaxy and background radio source are of entirely different redshifts. A few cases have been presented in the literature of \hi\ absorption due to intervening galaxies that belong to the same environment, \citep[for example, B2 1321+31, J1337+3152, 3C\,234, PKS 1649-062;][Mahony priv. comm., respectively]{Emonts_thesis, Srianand10, Pihlstrom_thesis}. However, the \hi\ absorption in all these cases is unresolved due to which a detailed study of the absorber has not been possible. Unlike these systems, for \target, it has been possible to characterise the absorber and, hence, to carry out a comparison with the samples of galaxies studied in \hi\ emission -- in the local universe and beyond.

At redshifts comparable to, or higher than that of, \target\ (i.e. $z \geq 0.1$), there are not many studies of \hi\ emission in galaxies \citep[e.g.][]{Catinella15, Hess19, Verheijen07, Gogate20}. The stellar masses of the galaxies included in such studies are comparable to that of our galaxy, while their \hi\ masses range from 2 $\times$ 10\pp{9} \Msun\ to a few times 10\pp{10} \Msun, considerably greater than the \hi\ mass of our galaxy. Similarly, their SFRs are between 0.3 \Msun\ yr\p{1} and 35 \Msun\ yr\p{1}, much higher than that in our case.

Thus, this galaxy, detected via \hi\ absorption, belongs to a population of galaxies which would be missed by the deep
\hi\ emission surveys. We note that this is likely to be a Malmquist bias since the emission studies are biased towards detecting the most HI-rich galaxies which is not the case with absorption studies. 

\begin{figure}
     \includegraphics[width=\linewidth]{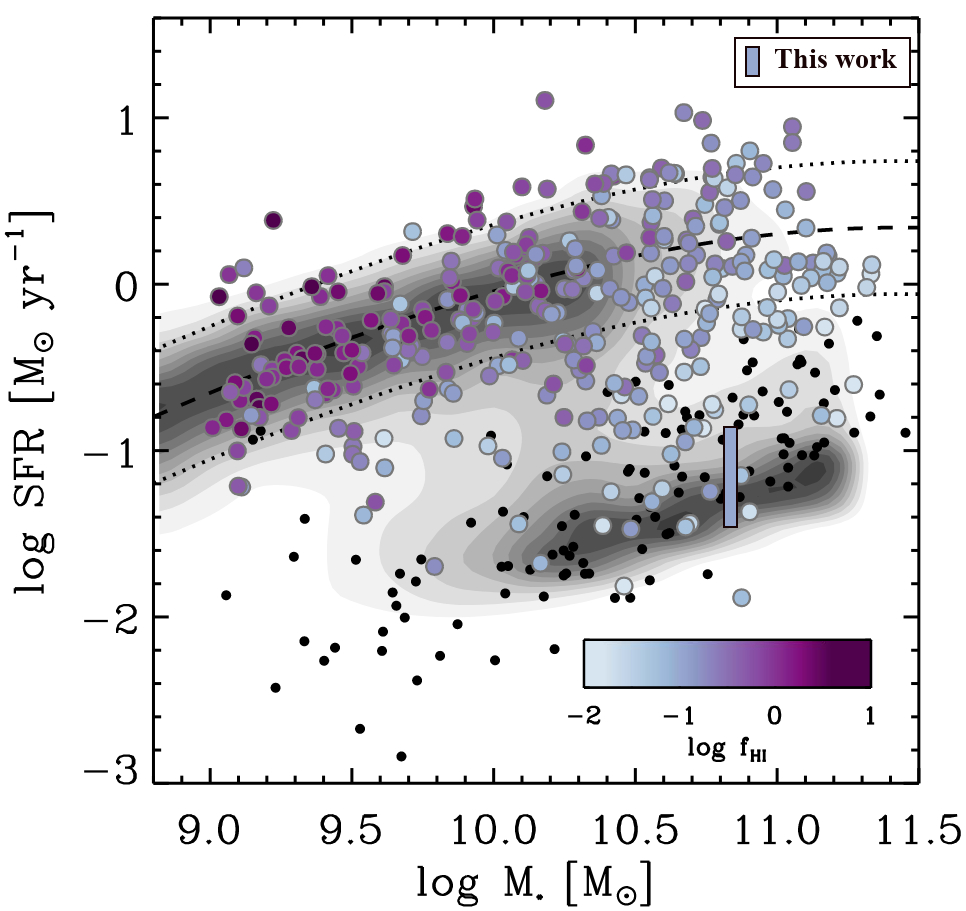}
    \caption{SFR$-$M$_{*}$ plane for xCOLD GASS sample from \citet{Saintonge17} with our \hi\ absorbing galaxy marked in blue rectangle (colour-coded on the atomic gas fraction). The black dots represent \hi\ non-detections in their sample while the grey contours represents the distribution of the SDSS galaxy population. The absorber galaxy is marked in a rectangle to represent the range of SFR derived; see Sect. \ref{sec:counterpart} for details. The f$_{\rm \hi}$, which corresponds to the ratio between \MHI\ and (\Mstar $+$ \MHI), for our \hi\ absorber galaxy ranges between --1.1 to --2.1 (in logscale) for a \tspin\ of 1000 K and 100 K respectively. For plotting, we consider a \tspin\ of 1000 K. \copyright AAS. Reproduced with permission.}
    \label{fig:saintonge17}
\end{figure}

Furthermore, we compare this galaxy with those studied in the local universe \citep[e.g. the xCOLD GASS sample by][]{Saintonge17}. Figure \ref{fig:saintonge17} shows the properties of the \hi\ absorber compared to galaxies of this sample. The rectangle indicates the range of SFR of the galaxy depending on whether the detected \halpha\ emission arises entirely from the SFR activity or not (see section \ref{sec:counterpart}). We find that the galaxy lies well below the star-forming main sequence. The upper limit to the atomic gas fraction obtained by considering a \tspin\ of 1000 K is comparable to the galaxies lying below the main sequence. However, a more reasonable estimate of f$\rm_{HI}$ obtained by assuming a \tspin\ of 100 K is lower even compared to the galaxies lying below the main sequence.

Finally, given that the absorber galaxy belongs to a different population compared to those studied at similar redshifts, it is also interesting to test whether this galaxy conforms with the baryonic Tully-Fisher relation (bTFR) \citep{McGaugh00}. To check the consistency with the baryonic Tully-Fisher relation, we need to determine the rotational velocity of the galaxy. We used 3D Barolo \citep{DiTeodoro15} to this end and found that the inclination of the galaxy is low ($\leq$ 20\pp{\circ}) and, hence, it is not quite possible to constrain the rotational velocity. However, assuming an inclination angle of 20\pp{\circ} results in a rotational velocity of \apx 120 \kmps, for which the galaxy is consistent with the bTFR \citep{McGaugh00}. For lower inclination angles, however, the galaxy is off the relation.

\subsection{Is the radio lobe interacting with the galaxy?}
\label{sec:interaction}

The detected \hi\ absorption is only \apx 50 \kmps\ blueshifted with respect to the systemic velocity of \target\ and, hence, the absorbing galaxy belongs to the same group of galaxies as \target. Because of this, it is of interest to investigate the possibility of an interaction between the galaxy and the radio AGN. Figure \ref{fig:optical_radio_halpha} shows the location of the galaxy with respect to the radio lobe. We find that the end of the radio lobe is not relaxed but, instead, protruded as if pinched at the corners. Intriguingly, the location of the galaxy and especially the majority of the \halpha\ emission arises, in projection, from the same region where the lobe appears pinched. It is quite possible that the radio lobe is interacting with the galaxy at this point and has acquired the observed morphology as a result. 

\halpha\ emission lends support to this possibility. Fig. \ref{fig:optical_radio_halpha} shows in detail the location of \halpha\ emission with respect to the radio lobe. We find that the emission has two components one arising from the central part of the galaxy and another located along the eastern boundary of the galaxy. It is very likely that these two components arise from two HII regions belonging to a spiral arm of the galaxy (see Fig. \ref{fig:GTCcompanion}). Interestingly, the \halpha\ blob along the eastern boundary of the galaxy also coincides with the boundary of the radio lobe in projection. Hence, we cannot rule out that an interaction with the radio lobe may be the cause of this blob. The radio lobe may have shock-ionised the gas in the galaxy, giving rise to the observed \halpha\ morphology. As has been seen in other cases, for example in Coma A \citep{Tadhunter00, Morganti02} and the Beetle galaxy \citep{Villar17}, interactions between radio plasma and gas are very much capable of producing such structures. In fact, in \target\ we observe such an effect within the AGN host galaxy as well as on the circumgalactic scales: in the radio host galaxy \target, the H$\alpha$ follows [OIII] emission that aligns with the inner radio jet (Fig. \ref{fig:overlayHST}), while blobs of H$\alpha$ emission are also found to align along the edge of the southern radio lobe in the circumgalactic environment between \target\ and the \hi\ companion (Fig. \ref{fig:Halpha}).

However, we would like to note that the morphology of the galaxy does not seem disturbed and the \hi\ kinematics of the galaxy appear regular. Hence, we suggest that if there is any interaction between the radio lobe and the galaxy, it is only present at a mild level. A further investigation of this scenario of interaction is possible by studying the spectral index of the southern radio lobe at the location of the galaxy. In the presence of an interaction of radio plasma with the interstellar medium of the galaxy, the electrons at those sites of interaction would be re-accelerated \eg{Harwood13}. Hence, we would expect the spectral index in those regions to be flatter compared to other regions of the lobe. A spectral index study of \target\ has been carried out by \citep{Lal07}, however, the spatial resolution of their study was not high enough to resolve the southern lobe sufficiently. Thus, a high-resolution spectral index study \citep[that may be carried out with, for example, LOFAR Two Meter Sky Survey][]{Shimwell17} would help confirm this scenario.

Although it is very unlikely, it might also be possible that the observed \halpha\ features are produced by the interaction of the radio lobe with the gas in the circumgalactic environment, which does not belong to the galaxy but overlaps with it only in projection. However, we do not have enough evidence to conclude either way based solely on the available data.

\subsection{Blind \hi\ absorption surveys in the SKA era}

As we see in the previous sections of this paper, resolved \hi\ absorption against an extended, bright-enough radio continuum can trace neutral hydrogen gas in galaxy populations beyond the local universe which would otherwise be missed by the current deep \hi\ emission surveys. With the SKA pathfinder facilities and the SKA itself, it may be possible to expand the number of such studies since they provide a large field of view and a simultaneous coverage of a large range of redshift, thus increasing the chance of detection of \hi\ absorption from intervening galaxies along the line of sight of an extended radio source in a single pointing.

Deep optical images of a large part of the sky are already available (e.g. SDSS, Pan-STARRS) and so are deep images of the radio sky at relatively high spatial resolution (e.g. VLASS, FIRST, LOFAR). A cross-match between the two would provide candidate systems where the extended radio emission overlaps, in projection, with a galaxy. Since the blind \hi\ absorption surveys presently planned with the SKA pathfinders  do cover a large area of the sky, it is possible to conduct a similar study of all these systems in a quasi-blind manner, without a preceding optical spectroscopic survey to ensure that the galaxy is in the foreground.

Apart from the chance alignment of extended, bright radio AGN and galaxies, high spatial resolution  is another crucial requirement for such studies. Even at relatively low redshifts, for example at $z \sim 0.2$, to be able to barely resolve a 50 kpc radio galaxy \citep[\apx 4 beam elements; typical size of a high-redshift radio galaxy, e.g.][]{Kanekar04}, the angular resolution needed is \apx 3$''$. The current and forthcoming blind \hi\ absorption surveys planned with the SKA-pathfinder facilities, namely, APERTIF \citep{Oosterloo09}, MeerKat \citep{Booth09} and ASKAP \citep{Johnston08}, will at best achieve an angular resolution of 6$''$ (by MeerKAT). Thus, further resolving the absorber galaxy which would only overlap with a small portion of the radio continuum may not be possible. However, even if unresolved, such studies will still be able to provide a census of low \hi\ mass galaxy population beyond the local universe, while the more detailed \hi\ mapping of galaxies in absorption may have to wait until SKA phase 1, whose sub-arcsecond angular resolution even at low frequencies (up to $z \sim 0.85$) and unprecedented sensitivity have the potential to expand the study of such systems to statistical samples.

\section{Summary}

In this paper, we present an \hi, \halpha,\ and optical continuum study of the \hi\ absorber against the radio galaxy \target. The peculiar radio morphology of \target, namely, the bright southern lobe, provides a favourable extended background continuum to resolve the foreground absorber. The absorber is only \apx 50 \kmps\ blueshifted from \target\ and, hence, it belongs to the same environment. 

Our resolved \hi\ absorption data obtained with the VLA in B array show that the absorber does not exhibit any sign of disturbance and has regular kinematics. With the help of our continuum and \halpha\ images from GTC, we find that the absorber is a disc galaxy located at the boundary of the southern radio lobe. This galaxy has a stellar mass of \apx 10\pp{10} \Msun\ and a maximum star-formation rate of \apx 0.15 \Msun\ yr\p{1}. We estimate the \hi\ mass of this galaxy to be \apx 3 $\times$ 10\pp{8} \Msun\ and, furthermore, we find that the \hi\ column density of the gas and star formation properties are representative of a gas-rich early-type galaxy off the star-forming main sequence.

We compare this result with the properties of galaxies detected in \hi\ in the local universe as well as at redshifts \apx 0.1. We find that for the given \hi\ and stellar mass, and the star-formation rate, it is consistent with the galaxies in the local universe. However, the deep \hi\ emission surveys at similar redshifts as our galaxy still have not detected galaxies of \hi\ masses as low as this. 

Since the absorber galaxy belongs to the same environment as \target, we also investigate the possibility of an interaction between the southern radio lobe and the galaxy. Interestingly, we find that in projection, the galaxy is located at the boundary where the radio lobe exhibits a protruded morphology. The \halpha\ emission from the galaxy shows two H\,\textsc{ii} regions, one of which coincides with the boundary of the radio lobe. Together with the fact that H$\alpha$ emission is also aligned with the radio continuum in the host galaxy of \target\ and at a few other locations along the southern radio lobe, this points to an interesting scenario where the radio AGN is directly interacting with a neighbouring galaxy.

Our study shows that in the case of a favourable alignment of a galaxy in front of an extended, bright-enough radio source, we can trace galaxy population in \hi\ that would otherwise be missed by deep \hi\ emission surveys. The deep blind \hi\ absorption surveys with the SKA-pathfinder facilities, in conjunction with the deep optical images available from the all-sky surveys, may be able to detect more systems of this kind, although they will fall short of the desired spatial resolution to resolve the absorption. The SKA phase 1 with sub-arcsecond resolution and high sensitivity out to high redshifts has the potential to extend the reach of such studies.

\begin{acknowledgements}
We thank the referee for useful comments. SM is grateful to Anqi Li and Thijs van der Hulst for valuable discussions. SM would also like to thank Scott Trager, Teymoor Saifollahi, and Pavel E Mancera Pi$\tilde{\rm n}$a for advice on analysing optical data. BE thanks Antonio Cabrera Lavers for advice on the GTC imaging. Part of the research leading to these results has received funding from the European Research Council under the European Union's Seventh Framework Programme (FP/2007-2013) / ERC Advanced Grant RADIOLIFE-320745. MVM acknowledges support from grant PGC2018-094671-BI00 (MCI/AEI/FEDER,UE). Her work was done under project No. MDM-2017-0737 Unidad de Excelencia “Mar\'\i a de Maeztu” Centro de Astrobiolog\'\i a (CSIC-INTA). The National Radio Astronomy Observatory is a facility of the National Science Foundation operated under cooperative agreement by Associated Universities, Inc. This work is based on observations carried out at the Observatorio Roque de los Muchachos (La Palma, Spain) with GTC (programme GTC48-17B). This research has made use of "Aladin sky atlas" developed at CDS, Strasbourg Observatory, France. We have also made use of the ``K-corrections calculator'' service available at http://kcor.sai.msu.ru/. The Pan-STARRS1 Surveys (PS1) have been made possible through contributions of the Institute for Astronomy, the University of Hawaii, the Pan-STARRS Project Office, the Max-Planck Society and its participating institutes, the Max Planck Institute for Astronomy, Heidelberg and the Max Planck Institute for Extraterrestrial Physics, Garching, The Johns Hopkins University, Durham University, the University of Edinburgh, Queen's University Belfast, the Harvard-Smithsonian Center for Astrophysics, the Las Cumbres Observatory Global Telescope Network Incorporated, the National Central University of Taiwan, the Space Telescope Science Institute, the National Aeronautics and Space Administration under Grant No. NNX08AR22G issued through the Planetary Science Division of the NASA Science Mission Directorate, the National Science Foundation under Grant No. AST-1238877, the University of Maryland, and Eotvos Lorand University (ELTE). IRAF is distributed by the National Optical Astronomy Observatories, which are operated by the Association of Universities for Research in Astronomy, Inc., under cooperative agreement with the National Science Foundation.
\end{acknowledgements}

%
   \bibliographystyle{aa} 
   \bibliography{AA_39114} 
%


\end{document}